# Classification and Evaluation the Privacy Preserving Data Mining Techniques by using a Data Modification–based Framework

MohammadReza Keyvanpour

Department of Computer Engineering
Al-Zahra University
Tehran, Iran
Keyvanpour@Alzahra.ac.ir

Somayyeh Seifi Moradi

Department of Computer Engineering
Islamic Azad University, Qazvin Branch
Tehran, Iran
S_seifi_moradi@yahoo.com

*Abstract*—In recent years, the data mining techniques have met a serious challenge due to the increased concerning and worries of the privacy, that is, protecting the privacy of the critical and sensitive data. Different techniques and algorithms have been already presented for Privacy Preserving data mining, which could be classified in three common approaches: Data modification approach, Data sanitization approach and Secure Multi-party Computation approach. This paper presents a Data modification–based Framework for classification and evaluation of the privacy preserving data mining techniques. Based on our framework the techniques are divided into two major groups, namely perturbation approach and anonymization approach. Also in proposed framework, eight functional criteria will be used to analyze and analogically assessment of the techniques in these two major groups. The proposed framework provides a good basis for more accurate comparison of the given techniques to privacy preserving data mining. In addition, this framework allows recognizing the overlapping amount for different approaches and identifying modern approaches in this field.

## I. INTRODUCTION

Although data mining can be valuable in many applications, it may cause to violation of privacy in case of no sufficient protection and abusing private data for other goals. The main factor of privacy beaching in data mining is *data misuse*. In fact, if the data consists of critical and private characteristics and/or this technique is abused, data mining can be hazardous for individuals and organizations. Therefore, it is necessary to prevent revealing not only the personal confidential information but also the knowledge, which is critical in a given field [1].

The principal attention to Privacy Preserving Data Mining (PPDM) is development of those algorithms, which - by protecting existed private data and knowledge in datasets and accessing the valid results of data mining-provide the possibility to share the critical and private data for analytical aims.

There are two general scenarios in Privacy Preserving Data Mining: the Multi-party collaborations scenario and Data publishing scenario. In the former, the collection of data is distributed between two or more sites, each one owns a part of the private data and these sites collaborate to compute a data mining algorithm on the union of their databases without revealing the data at their individual sites and the results of data mining will only be revealed. The major approach for this scenario is the Secure Multi-party Computation.

In Data publishing scenario the owners or data providers are publishing or sharing their data to acquire data mining results and /or joining the data mining process. In this scenario, as shown in figure (1), the privacy preservation techniques are applied during the data integration or before sending data to the data miner.

Principal approaches in this scenario based on the goal of privacy preservation- classified in two categories: Data modification and Data sanitization.

Data sanitization approaches aim to hide the critical rules and patterns existed in dataset. However, the Data modification approaches are hiding critical data and aiming to acquire valid results of data mining while private data cannot be reached directly and precisely. In these techniques, major concerns are to maximize the quality of the released data, data mining results accuracy and protecting the data privacy as well.

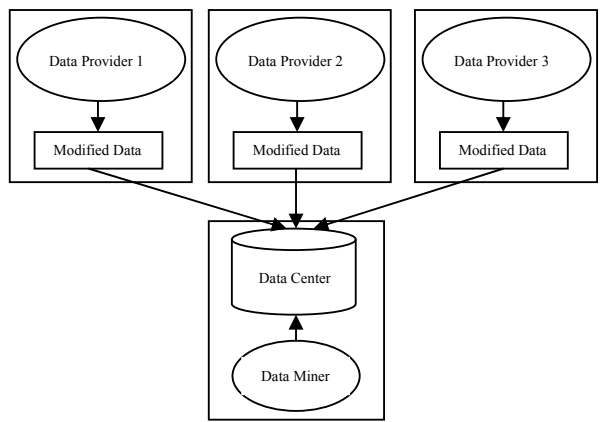

Figure 1. PPDM based on Data publishing scenario

This paper attempts to provide a good basis for classification and more accurate evaluation of data modification–based techniques for Privacy Preserving Data Mining. The rest of the paper is organized as follows. In section 2, the recommended classification framework for data modification techniques in PPDM will be given and then we introduce these techniques. In section 3 we propose the evaluation framework and analyze these techniques under this framework and finally, in section 4, the paper will be finalized by conclusion as well.

## II. CLASSIFICATION FRAMEWORK

Data modification techniques study and review for PPDM shows that these techniques can be classified in two principle groups of perturbation-based and anonymization-based techniques according to how the protection of privacy. The recommended classification framework is shown in figure (2).

Anonymization techniques are preventing from recognizing the critical data's characters and identity to preserve the privacy while perturbation approach modify a part of data or the whole dataset by means of determined techniques and in a manner to save the particular properties, which are meaningful and significant for creating data mining models.The current techniques in perturbation approach are classified in two categories based on how they perturb datasets and particular Properties that will be preserved in data: Value-based Perturbation and Multi-Dimensional Perturbation. In the Value-based Perturbation the purpose is to preserve statistical characteristics and columns distribution while Multi-Dimensional Perturbation aims to hold Multi-Dimensional information.

### A. Anonymization Techniques

First natural solution to publish raw-critical data with privacy preserving is de-identification in which the raw-critical dataset is spread after removing the key identities of the records. But, in combination with an external database, there might be some other attributes to be used for identifying the personal details, called "Quasi-Identifier" (QI). To solve re-identifying problem, anonymization approach was brought up, in which the values of the QI attributes become modified so that they no longer uniquely represent individuals.

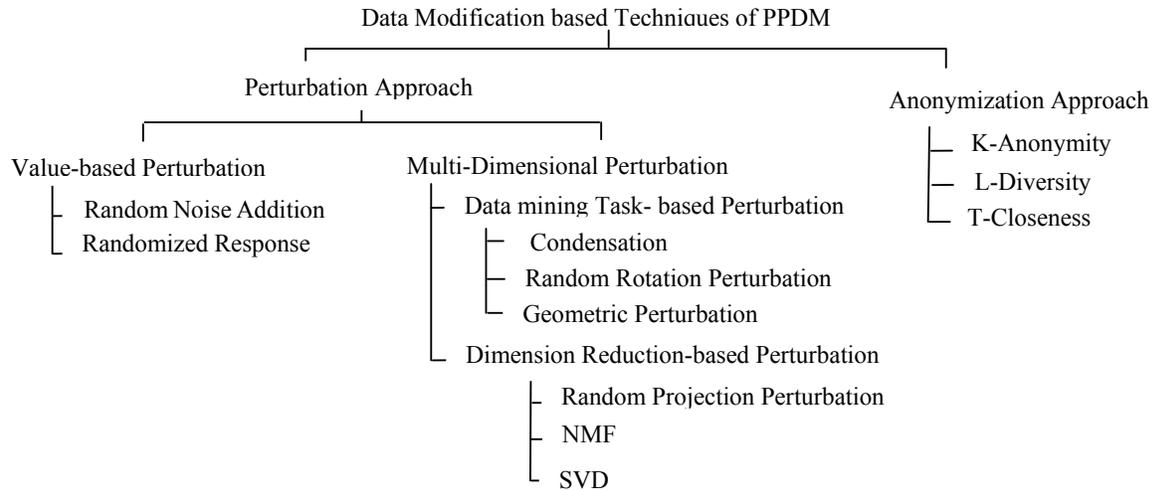

Figure 2. Data Modification-based framework for classification the PPDM Techniques

### 1) K-anonymity Technique

In this technique each record within an anonymized table must be indistinguishable with at least k-1 other record within the dataset, with respect to a set of QI attributes. In particular, a table is K-anonymous if the QI attributes values of each record are identical to those of at least k-1 other records. To achieve the K-anonymity requirement, generalization or suppression could be used [2, 3].

Nevertheless, this technique consists of some limitations. First, it is very difficult for a database owner to determine which of the attributes are or are not available in external tables. Second, this model considers a certain type of attack (Linkage attack) and cannot preserve sufficiently the sensitive attributes against Homogeneity attack (similarity of the sensitive attributes values in an anonymized group) and Background Knowledge attack (awareness about the relationship between sensitive and QI attributes).

### 2) L-diversity Technique

This technique [4] was proposed to solve the Homogeneity attack of K-anonymity technique that emphasizes not only on saving the minimum size of $K$ group but also considers saving the variety of the sensitive attributes of each group. In this technique, every anonymized group must hold at least $l$ well-represented values for each sensitive attribute. However, this technique has some shortcomings too: e.g. it might be unnecessary and difficult to achieve that. On the other hand, this technique is insufficient to prevent attribute disclosure, Such as Similarity Attack. In fact, if the sensitive attribute values in an anonymized group are distinct but semantically the same, the adversary can learn important information.

### 3) T-Closeness Technique

As it was discussed, L-diversity technique's deficiency is that it treats all values of a given attribute in a same way regardless of its distribution in the data, while this rarely happens for real datasets in which the values of the sensitive attributes are not probably in the same sensitivity level; in this way, the precise values of sensitive attributes might be inferred by the use of Background Knowledge Attack. In T-Closeness Technique [5] the distance between the distribution of a sensitive attribute in an anonymized group and its distribution in the whole table should not be more than $t$ threshold. The main challenge of this task is how to show the distance criterion which can reflect the semantic gap between the quantities.

Overall, anonymization techniques are simple and their main advantage is scalability toward privacy preservation (choosing greater $K$); however, they have an inherent weakness that they cannot always prevent the records' critical values deduction against attacks efficiently. Moreover, it has been shown that optimal anonymization is an "NP-Hard" problem [6]. Furthermore, this technique is not even effective with increasing dimensionality, since the data can typically be combined with either public or background information to reveal the identity of the underlying record owners [7].

### B. Value-based Perturbation Techniques

The main idea of this approach is to add random noise to the data values. This approach is actually, based on this fact that some data mining problems do not need the individual records necessarily and they just need their distribution. Since the perturbing distribution is known, they can reach data mining goals by reconstructing their required aggregate distributions. However, due to reconstructing each data dimension's distribution independently, they have the inherent disadvantage of missing the implicit information available in multi-dimensional records and on the other hand it is required to develop new distribution-based data mining algorithms.

*1)  Random Noise Addition Technique*

This technique is described in [8] as follows: Consider $n$ original data $X_1, X_2 \ldots, X_N$, where $X_i$ are variables following the same independent and identical distribution (i.i.d). The distribution function of $X_i$ is denoted as $F_X$ , $n$ random variables $Y_1, Y_2 \ldots, Y_N$ are generated to hide the real values of $X_i$ by perturbation. Similarly, $Y_i$ are *i.i.d* variables. Disturbed data will be generated as follows:

$$w_1, \ldots, w_n \ where \ w_i = X_i + Y_i \ \ i = 1, \ldots, n \qquad (1)$$

It is also assumed that the added noise variance is large enough to let an accurate estimation of main data values take place. Then, according to the perturbed dataset $w_1, \ldots, w_n$, known distributional function $F_Y$ and using a reconstruction procedure based on Bayes rule, the density function $f_X^{'}$ will be estimated by Equation (2).

$$f_X^{'}(a) = \frac{1}{n} \sum_{i=1}^{n} \frac{\int_{-\infty}^{a} f_Y\left(w_i - a\right) f_X\left(a\right)}{\int_{-\infty}^{+\infty} f_Y\left(w_i - z\right) f_X\left(z\right) dz} \qquad (2)$$

Although $f_X$ isn't really known, we can estimate it by using the normal distribution as the initial estimate and iteratively refine this estimate by applying Equation (2).

In [9], to minimize the information loss of this technique and improve the reconstruction procedure, a new distribution reconstruction algorithm called *Maximizing Algorithm for the Expectation of Mathematics* (EM) is represented. In [8] a new decision-tree algorithm is developed according to this technique. This technique is also used in privacy preserving association rule mining [10, 11].

However, in [12] it is presented that privacy breaches as one of the major problems with the random noise addition technique and observed that the spectral properties of the randomized data can be utilized to separate noise from the private data. The filtering algorithms based on random matrix theory are used to approximately reconstruct the private data from the perturbed data. Thus, establishing a balance between Privacy preservation and accuracy of data mining result is hard because more we want privacy preservation, more we should lose information.

*2)  Randomized Responses Technique*

The main idea for this technique [13] is to scramble data so that the data collector cannot express, with a probabilities better than of the defined threshold, whether the data sent back by the respondent is correct or not. There are two models in this technique: *Related–Question* and *Unrelated- Question* models. In the former, the interviewer asks every respondent a couple of questions related together, of each the reply is opposite the other one. For example, the questions can be as follows:

1)  I have the sensitive attribute A.
2)  I do not have the sensitive attribute A.

The respondent will answer randomly and with $\theta$ probability to the first question and with *1- θ* to the second question. Although the interviewer finds out the answers (yes or no), he does not know which question has been answered by the respondent; hence, the respondent's privacy preserving will be saved. The collector uses the following equations in order to estimate the percentage of the people who have characteristic A:

$$P^*\left(A = yes\right) = P\left(A = yes\right).\theta + P\left(A = no\right).\left(1 - \theta\right) \qquad (3)$$

$$P^*\left(A = no\right) = P\left(A = no\right).\theta + P\left(A = yes\right).\left(1 - \theta\right) \qquad (4)$$

Where $P^*(A = yes)$ (or $P^*(A = no)$) is the ratio of the "*yes*" (or "*no*") replies which are acquired via survey data and $P(A = yes)$ (or $P(A = no)$) are estimated ratios of the "*yes*" (or "*no*") answers to sensitive questions. The purpose is to gain $P(A = yes)$ $P(A = no)$.

Although in this method, information from each individual user is scrambled, if the number of users is significantly large, the aggregate information of these users can be estimated with decent accuracy. Randomized response technique used to provide information with response model, so are used for processing categorical data. Note that the technique can be extended to multi-dimensional, i.e., the techniques are applied to several dimensions altogether [23].

### C. Data Mining Task-based Perturbation Techniques

The purpose of these techniques is to modify the original data so that the properties preserved in perturbed dataset to be task specific information data mining tasks and even a particular model. Thus, it is possible to preserve the privacy without missing any particular information of data mining tasks and make a more suitable balance between privacy and data mining results accuracy. Furthermore, in these techniques, data mining algorithms can be applied directly and without developing new data mining algorithms on the perturbed dataset.

#### 1) Condensation Technique

The purpose of this technique is to modify the original dataset into anonymized datasets so that this anonymized dataset preserves the covariance matrix for multiple columns. In this technique first the data will be condensed into groups with pre-defined size $K$, and a series of statistical information related to the mean and correlations across the different dimensions will be preserved for each group of records. In the server, this statistical information is used to generate anonymized data with similar statistical characteristics to the original dataset. This technique has been used to create simple classifier for the K Nearest Neighbor (KNN) [14].

However in [15] it is presented that this technique is weak in protecting the private data. The KNN-based data groups result in some serious conflicts between preserving covariance information and preserving privacy.

#### 2) Random Rotation Perturbation Technique

The main idea is as if the original dataset with $d$ columns and $N$ records represented as $X_{d \times n}$, the rotation perturbation of the dataset X will be defined as $G(X) = RX$. Where $R_{d \times d}$ is a random rotation orthonormal matrix.

A key feature of rotation transformation is preserving the Euclidean distance, inner product and geometric shape hyper in a multi-dimensional space. Also, kernel methods, SVM classifiers with certain kernels and hyper plane-based classifiers, are *invariant* to rotation perturbation, i.e. if trained and tested with rotation perturbed data, will have similar model accuracy to that trained and tested with the original data. [15].

But researches show that having previous knowledge, the random rotation perturbation may become involved in privacy violations against different attacks including Independent Component Analysis (ICA), attack to rotation center and distance-inference attack [16, 17].

#### 3) Geometric Perturbation Technique

This perturbation technique is a combination of Rotation, Translation and Noise addition perturbation techniques. The additional components $\psi$ and $\Delta$ are used to address the weakness of rotation perturbation while still preserving the data quality for classification modeling. Concretely, the random translation matrix addresses the attack to rotation center and adds additional difficulty to ICA-based attacks and the noise addition addresses the distance-inference attack.

If the matrix $X_{d \times n}$ indicates original dataset with $d$ columns and $N$ records, $R_{d \times d}$ be a orthonormal random matrix, $\psi$ be a translation random matrix and $\Delta_{d \times n}$ be a random noise matrix, where each element is Independently and Identically Distributed (iid) variable like Gaussian distribution N(0,σ2), the geometrical perturbation will be defined as following [17]:

$$G(X) = RX + \psi + \Delta \qquad (5)$$

Definition: let $t_{d \times 1}$ represent a random vector and $l_{1 \times n}$ a vector of "N" '1's. Matrix $\psi$ is a translation matrix if $\psi = [t, t, \ldots, t]_{d \times n}$, i.e. $\psi_{d \times n} = t_{d \times 1} l_{n \times 1}^T$.

This perturbation technique is invariant against geometrical modification and is fixed for Kernel, SVM and linear classifiers. Geometrical perturbation technique also has, rather than Rotation perturbation and condensation, high-great Privacy Preserving guarantees.

### D. Dimension Reduction-based Perturbation Techniques

The main purpose of these techniques is to obtain a compact representation with reduced-rank to the original dataset while preserving dominant data patterns. These techniques also guarantee that both the dimensionality and the exact value of each element of the original data are kept confidential.

#### 1) Random Projection Perturbation Technique

Random projection [16] refers to the technique of projecting a set of data points from a high-dimensional space to a randomly chosen lower-dimensional subspace. If the matrix $X_{m \times n}$ (or $Y_{m \times n}$) indicates original dataset, $R_{n \times k}$ $(k < n)$ ( or $R'_{k \times m}$ $(k < m)$ ) be a random matrix such that each entry $r_{i \times j}$ of $R$ (or $R'$) is independent and identically chosen from some unknown distribution with mean zero and variance $\sigma_r^2$, the Column-wise Projection $G(X)$ and Row-wise Projection $G(Y)$ will be defined as below:

$$G(X) = \frac{1}{\sqrt{k}\,\sigma_r} XR \;,\; G(Y) = \frac{1}{\sqrt{k}\,\sigma_r} R'Y \qquad (6)$$

The key idea of random projection arises from the *Johnson-Lindenstrauss Lemma* [18]. According to this lemma, it is possible to maintain distance-related statistical properties simultaneously with dimension reduction for a dataset. Therefore, this perturbation technique can be used for different data mining tasks like including inner product/Euclidean distance estimation, correlation matrix computation, clustering, outlier detection, linear classification, etc.

However, this technique can hardly preserve the distance and inner product during the modification in comparison with geometric and random rotation techniques. It has been also clarified that having previous knowledge about this perturbation technique may be caught into privacy breach against the attacks [17].

#### 2) Singular Value Decomposition (SVD) Technique

The SVD [19] is a well-known method of dimension reduction in data mining process. If $A_{n \times m}$ matrix indicates the original dataset, then SVD for data matrix $A$ will be as:

$$A = U \Sigma V^T \qquad (7)$$

Where $U_{n \times n}$ is an orthonormal matrix, $V^T_{m \times m}$ is an orthonormal matrix and $\Sigma_{n \times m}$ is a diagonal matrix whose nonnegative diagonal entries are the singular values in a descending order,

$$\Sigma = \mathrm{diag}\big[\sigma_1, \sigma_2, \ldots, \sigma_s\big] \quad \big(s = \min\{m, n\}\big) \qquad (8)$$

Due to the arrangement of the singular values in the matrix $\Sigma$ (in a descending order), the SVD transformation has the property that the maximum variation among the objects is captured in the first dimension. Similarly, much of the remaining variations are captured in the second dimension, and so on. Thus, a transformed matrix with a much lower dimension can be constructed to represent the structure of the original matrix faithfully, defined as below:

$$A_k = U_k \Sigma_k V_k^T \; where \; k \ll \min(n, m) \qquad (9)$$

Where $U_k$ includes first $k$ columns of U, $\Sigma_k$ includes first $k$ non-zero singular values, and $V_k^T$ includes the first $k$ rows of $V^T$ . The rank of the matrix $A_k$ is $k$.

In data mining applications the omitted part $E_k = A - A_k$ could be considered as the noise in the original dataset. Hence, in many of conditions, mining over the reduced dataset $A_k$ can produce better results than mining on the original dataset. When this technique is applied for Privacy Preserving purposes, the distorted dataset $A_k$ can protect the privacy and simultaneously, keeps the utility of the original data as it can faithfully represent the original data structure. [20].

### 3) Non-negative Matrix Factorization (NMF) Technique

NMF [21] is a matrix factorization method to obtain a representation of data using nonnegative constraints. Considering a $n \times m$ nonnegative matrix dataset A with $A_{ij} \geq 0$ and a pre-specified positive integer $k \leq \min\{n, m\}$, nonnegative matrix factorization (NMF) finds two non-negative matrixes $W \in R^{n \times k}$ with $W_{ij} \geq 0$ and $H \in R^{k \times m}$ with $H_{ij} \geq 0$, such that $A \approx WH$ and the objective function is minimized:

$$f(W, H) = \frac{1}{2} \|A - WH\|_F^2 \qquad (10)$$

$W$ and $H$ matrixes have many mostly-desired Properties in data mining applications. In [22] the NMF technique is used for Privacy Preserving in Data mining applications.

Recently, accordance with classification algorithms, it has been shown that SVD and NMF provide much higher degree of data distortion than the standard data distortion techniques based on adding uniformly distributed noise or normally distributed noise. Moreover, these techniques consist of a high-level accuracy in data mining results as well as high-level privacy preserving.

## III. EVALUATION FRAMEWORK

The evaluation framework recommended for assessing and evaluating data modification-based techniques, is in accordance with the following eight criteria:

- *Privacy Loss*: is defined as difficulty level in estimating the original values from the perturbed data values.

- *Information Loss*: is defined based on the amount of important data information, which needs to be saved after perturbation for data mining purposes.

- *Data Mining Task*: is defined based on a data mining task, which contains the possibility to mine it, after applying the privacy preserving techniques.

- *Modifying the Data Mining Algorithms*: based on the needs, notifies the change for the existed data mining algorithms, in order to mine over the modified dataset.

- *Preserved Property*: that is, data information, which was already saved after applying the privacy preservation techniques.

- *Data Type:* it points out the types of data, which could be numerical, binary, or categorical.

- *Indistinguishability Level*: it is in accordance with level of indistinguishability of different records of the original dataset.

- *Data Dimension:* it is defined based on the purpose of PPDM technique for preserve Dimensional information, which could be single-Dimensional or Multi-Dimensional.

The techniques based on data modification in Privacy Preserving Data Mining were already analyzed and assessed based on the above-mentioned criteria, as shown in table (1).

TABLE I.        ASSESSMENT FRAMEWORK FOR THE TECHNIQUES BASED ON DATA MODIFICATION IN PPDM

| Comparison Criteria | | Data Modification based Techniques of PPDM | | | | | | | | | |
|---|---|---|---|---|---|---|---|---|---|---|---|
| | | Anonymization | | | Value-based Perturbation | | Multi Dimensional Perturbation | | | | | |
| | | | | | | | Data Mining Task-based Perturbation | | | Dimensional Reduction-based Perturbation | | |
| | | k-an | L-div | T-clo | Noise Addition | Randomized Response | Condensation | Random Rotation | Geometric | Random Projection | NMF | SVD |
| Privacy Loss | | Average | | | Average | Average | Low | Low | Very Low | Very Low | Very Low | |
| Information Loss | | Low | | | Low | Low | Very Low | Very Low | Very Low | Very Low | Very Low | |
| Modifying DM Algorithms | | No | | | Yes | Yes | No | No | No | No | No | |
| Data Mining Task | Asso | | | | √ | | | | | | | |
| | Class | √ | | | √ | √ | √ | √ | √ | √ | √ | |
| | Clus | | | | | | | | | √ | | |
| Data Dimension | | Multi-Dimensional | | | single-Dimensional | Single Dimensional | Multi Dimensional | Multi Dimensional | Multi Dimensional | Multi Dimensional | Multi Dimensional | |
| Preserved property | | - | | | Values distribution | Values distribution | Covariance structure | Geometrical characteristic | Geometrical characteristic | Corelation between dimension | Corelation between dimension | |
| Data type | | - | | | - | Categorical | numerical | numerical | numerical | numerical | numerical | |
| Indistinguishabil_ity Level | | k | | | - | - | k | - | - | - | - | |

## IV.    CONCLUSION

In this paper, a data modification-based framework was presented for classification and evaluation the Privacy Preserving Data Mining techniques. At first, these techniques classified into two classes of anonymization and perturbation approaches and after analyzing each approach, their significant characteristics were given. The main challenge of anonymization approach was insufficient protection of critical values deduction against different attacks and being NP-Hard optimal anonymization. Instead, in perturbation process although it has a great efficiency from computation cost point of view, creating a suitable and stable balance between privacy and data mining results accuracy in that, is difficult.

Hence how to create a better balance between privacy and accuracy, how to further improve the algorithms efficiency and privacy preserving generality in different types and different Data mining tasks are some of the research aspects in the future.